\documentclass[prl,twocolumn,showkeys,showpacs,preprintnumbers,amssymb,amsmath,superscriptaddress]{revtex4}

\usepackage{graphicx}
\usepackage{bm}
\usepackage{dcolumn}
\begin{document}

\title{Dislocation formation from a surface step in semiconductors: an {\it ab initio} study}

\date{\today}

\author{J. Godet}

\altaffiliation{Current address: EPFL, IRRMA, B\^at. PPH 332
(CRPP), Station 13, CH - 1015 Lausanne, Switzerland}

\author{S. Brochard}

\author{L. Pizzagalli}
\email{Laurent.Pizzagalli@univ-poitiers.fr}

\author{P. Beauchamp}

\affiliation{Laboratoire de M\'etallurgie Physique, UMR 6630, Universit\'e
de Poitiers,  B.P. 30179, 86962 Futuroscope Chasseneuil Cedex, France}

\author{J. M. Soler}

\affiliation{Dep. de F\'isica de la Materia Condensada, Universidad Aut\'onoma de Madrid, E-28049 Madrid, Spain}

\begin{abstract}
The role of a simple surface defect, such as a step, for relaxing the stress
applied to a semiconductor, has been investigated by means of large scale first
principles calculations. Our results indicate that the step is the privileged
site for initiating plasticity, with the formation and glide of 60$^\circ$
dislocations for both tensile and compressive deformations. We have also
examined the effect of surface and step termination on the plastic mechanisms.

\end{abstract}



\preprint{report number}

\maketitle

The plasticity of semiconductors has been extensively studied for the last
decades in both fundamental and applied research, leading to significant
progresses in the understanding of the key mechanisms involved. Several issues
remain unsolved, however, one of the most essential being the formation of
dislocations in nanostructured semiconductors such as nano-grained materials,
or nano-layers in heteroepitaxy, systems extensively used in devices. While in
bulk materials the few native dislocations are able to multiply via Frank-Read
type mechanisms to ensure plasticity, the situation is different in
nanostructured materials where dimensions are too small to allow dislocation
multiplication \cite{Mou99PMA}. The presence of dislocations in these materials
appears to be more controlled by nucleation than by multiplication processes.
It has been proposed that surfaces and interfaces, which become prominent for
small dimensions, play a major role. Several observations support this
assumption, especially for strained layers and misfit dislocations at
interfaces \cite{Dun97JMSME,Jai97PMA,Wu01PMA}. The
formation at surfaces is also relevant where large stresses exist, like near a
crack \cite{Wu98PML,Sca99PSS,Rab01MSE,Pir01PMA,Pir99JMR}.

Since {\em in situ} experimental observations of dislocation nucleation is not
yet possible due to the very small dimensions and short observation timescales,
the formation of dislocations at surfaces has been mainly investigated
theoretically, particularly with continuum models and elasticity theory
\cite{Fra50Sym, Kam90JAP, Bel93PSS}. However, in these approaches, the
predicted activation energy is very large, in disagreement with experiments. It
has been proposed that surface defects, such as steps, help the formation by
lowering the activation energy. This is supported by experimental facts in the
context of dislocation nucleation at or near crack fronts, with dislocation
sources located on the cleavage surface and coinciding with cleavage ledges
\cite{Geo93MSE, Mic99NATO, Gal01PMA, Arg01SM2}. In addition, it has been shown
that, in a stressed solid, a surface step is a source of local stress
concentration \cite{Mar63INT,Smi68IJES,Cul96JCG}, although not as efficient as
a crack tip. Therefore, a number of continuum models have been developed,
taking into account the energy gain associated to the step elimination in the
process of dislocation nucleation
\cite{Jag91,Bel95Sym,Zou96JAP,Jun97PML}. Atomistic calculations
have also been performed for characterizing the energetics, the processes
involved, and the role of surface defects
\cite{Nan93AMM,Ich95,Jua96PML,Zim96EMRS,Asl98PML,Don98JAP,Bro00}.

These studies led to a better knowledge of the dislocation formation from
surface steps or cleavage ledges, but we are still far from a complete
understanding of the phenomenon. Furthermore, studies were mostly focussed on
ductile materials, such as metals, using empirical potentials. In
contrast, there is a certain lack of knowledge regarding semiconductors, for
which a different behavior is expected. Also, the role of the step has
not been identified. Another point of concern is the interatomic
potential for modeling dislocations. While
sufficiently reliable potentials have been developed recently for some metals,
the same is not yet true for a model semiconductor such as silicon
\cite{Piz03PMA,God03JPCM}.

In this work, we report investigations of the dislocation formation from a
surface step in a stressed semiconductor, here silicon. An {\em ab initio}
approach has been employed because of the insufficient reliability of empirical
potentials for modeling the rearrangement of atomic bonds that occurs during
the formation and propagation of dislocations. Our calculations clearly indicate
that a step is a privileged site for initiating plastic deformation. Indeed, a
60$^\circ$ dislocation forms from the step both in compression and traction
simulations. We also show the importance of surface and step termination,
finding that the dislocation forms below the surface when the surface and the
steps are passivated.

\begin{figure*}[t]
\includegraphics[width=16cm]{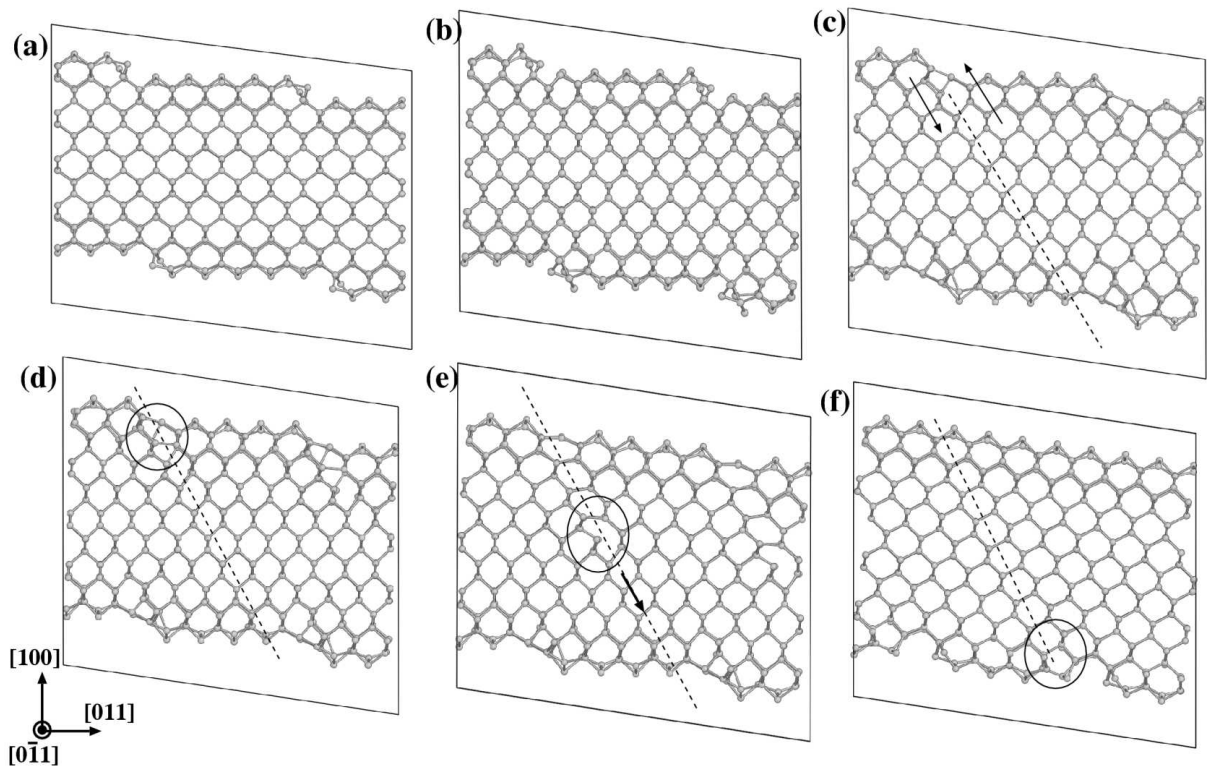}
\caption{Different stages of the compression process for a 196 atoms unit cell:
(a) Unstrained system, (b) strain of $-$10.4\%, (c) strain of $-$11.5\%,
(d-f) strain of $-$13.6\%. The dashed line shows the shuffle plane where the
dislocation glides. Two unit cells are drawn for clarity.}\label{SZ196}
\end{figure*}

Our calculations use density functional theory, in the local density
approximation, and norm-conserving pseudopotentials \cite{Tro91PRB}. We have
used the SIESTA method \cite{Ord96PRB,Sol02JPCM} with a basis set of atomic
orbitals. In order to simulate the largest possible systems, we searched
carefully the least expensive basis, able to accurately model the dislocation
formation. We used a minimal basis of optimized orbitals with a maximum range
of 6~\AA\ . The grid for numerical integration in real space has an energy
cutoff of 150~Ry. Two or four special k-points \cite{Mon76PRB} were used for
the Brillouin zone sampling, depending on the size of the system. These
parameters lead to accurate values for the lattice parameter and elastic
coefficients. Another test was the energy variation when bulk silicon is
strained along the \{111\} dense planes in the $\langle110\rangle$ direction,
the direction of Burgers vector dislocations in the diamond cubic structure
\cite{Hir82WIL}. The calculated shear strength is 28\% larger than the one obtained
with another basis, more accurate but much more expensive, but it is reached for
the same shear strain \cite{God03JPCM}.
Therefore we expect that the minimal basis set is adequate to study the mechanisms of
dislocation formation.

A typical model used in our simulations is shown in
Figure~\ref{SZ196}-a. The slab includes two (100) surfaces, with a
p(2$\times$1) reconstruction of asymmetric dimers. Steps lying
along the [0$\bar1$1] directions, which correspond to the
intersection of \{111\} slip planes and the (100) surface, are
placed on both surfaces. We have used double layer steps, as
formed by the emergence of a perfect dislocation at the surfaces
\cite{God04PRB}. Tilted periodic boundary conditions are applied
normal to the step direction, in order to have only one step on
each surface. For the periodic boundary conditions along the step
line direction [0$\bar1$1], four atomic planes are considered,
allowing the p(2$\times$1) reconstruction. The total number
of atoms in the system ranged from 124 to 508, depending on the
number of layers along [100] and [011], the normals to the step
line. We impose an increasing uniaxial stress contained in the
surface and making an angle $\alpha$ with the step normal [011],
by applying a strain.  We have shown previously from Schmid factor
analysis and empirical potential calculations that
$\alpha=22.5^\circ$ was the easiest orientation to form
dislocations \cite{God04PRB}, then unless explicitly stated the
results presented here are for this stress orientation. After
each stress increment of 1.5 GPa ($\sim 1$\% deformation), the
atomic positions were relaxed with a conjugate gradient algorithm
until the atomic forces were lower than 0.04~eV/\AA. The relaxed
configurations are then relevant for a crystal at 0~K.

The short period along the step direction allows only for the
formation of straight defects, no half-loop dislocations or even
kinks can form. However, there is no {\em a priori} restriction on
the type of straight dislocation, either perfect or partial, in
the shuffle set or the glide set \cite{shuffleglide}. These simulations are
representative of the low temperature behavior.

\begin{figure*}[t]
\includegraphics[width=16cm]{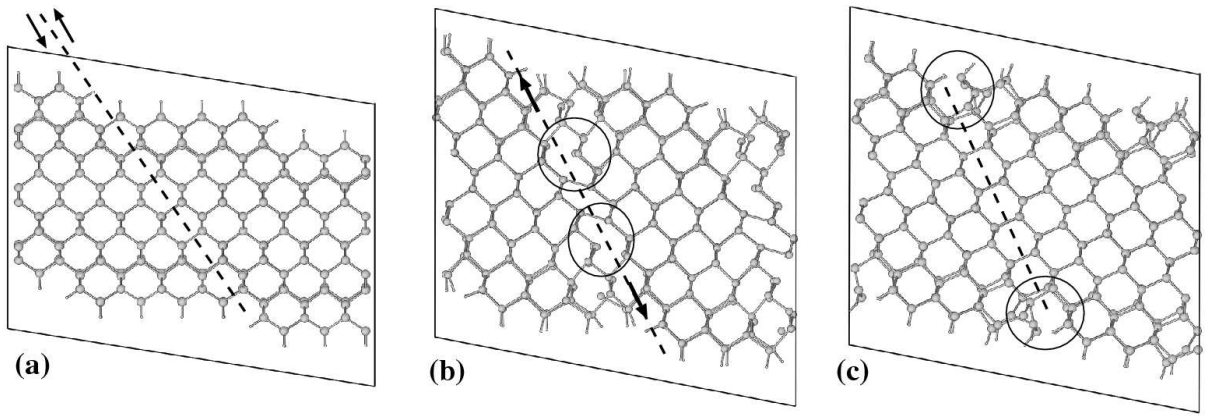}
\caption{Different stages of the compression process for a 128 atoms system,
with surfaces passivated with hydrogen atoms (white balls). (a) Unstrained
system projected along [0$\bar1$1], (b-c) $-16.7$\% of strain, formation and
glide of a dislocation dipole in bulk associated with the shear strain of the
(111) shuffle set plane crossing the steps. Two unit cells are drawn for
clarity. }\label{sat_h}
\end{figure*}

For a $D_B$ non rebonded step \cite{Cha87PRL},
we found that the 196 atoms system behaves elastically up
to a compressive strain of $-13.6$\%, which corresponds to a
linear stress of 19.5~GPa$=0.16G$, $G$ being the shear
modulus (Fig.~\ref{SZ196}). This is lower than
the theoretical shear strength calculated using the same method,
$0.21G$. At $-11.5$\% there is first a formation of
bonds between the atoms of the step edges and those of the lower
terraces (Fig.~\ref{SZ196}-c). Then at $-13.6$\%, as relaxation
continues, a defect is formed which, through successive breaking
and formation of bonds, glides from the top to the bottom surface
(Fig.~\ref{SZ196}-d,e). Finally, a new step forms on the bottom
surface, leaving a now perfect top surface (Fig.~\ref{SZ196}-f).
An analysis revealed that a perfect 60$^\circ$ dislocation formed
and slipped on the \{111\} shuffle plane that
passes through the step edge of the top surface. The process
removes almost all of the applied stress, with only 1.5~GPa
remaining. This stress value corresponds roughly to the deformation increment of 1\%. It is likely that 
this remaining stress could be reduced by using smaller increments. 

Qualitatively, we found quite similar results for traction.
Plasticity occurred at 21.9\% (31.5~GPa$=0.26G$) with several
atomic rearrangements in the vicinity of one step, leading to a
locally disordered crystal. Then, a 60$^\circ$ dislocation formed
from this area and slipped in a shuffle plane toward the opposite
surface. Therefore, our results clearly indicate that surface
steps facilitate the formation of dislocations in covalent
materials. As a consequence, the elastic limit is lowered,
compared to the bulk or to the perfect surface. The step breaks
the surface symmetry and facilitates the nucleation of plastic
events in its vicinity. This effect may be attributed either to a stress concentration or 
to a local reduction of elastic constants. A deeper analysis is difficult, since both contributions are intimately linked.

In order to check whether the step geometry has an effect on the
kind of formed dislocation, a calculation with a $D_B$ rebonded
step has also been carried out. This step can be thought as formed
by the emergence at the surface of a perfect dislocation located
in a glide set plane, followed by surface reconstruction.
In that case a 60$^\circ$ perfect dislocation is
also nucleated in the shuffle set plane near the step, but for a
larger strain (16.7\%), as already observed with classical
empirical potentials calculations for $\alpha=0^\circ$
\cite{God02SM}. Thus this suggests that for all step geometries, 
a 60$^\circ$ perfect dislocation is nucleated in a
shuffle set plane for this stress orientation. Such a result may be surprising, since it 
is known that plasticity of silicon deformed at high temperature is governed by dissociated dislocations located in the glide set. In that case, dislocations move 
with the formation and migration of double kinks. 
However, it has been shown that in the low temperature/high stress regime, the plasticity is dominated by perfect dislocations located in shuffle
planes \cite{Rab01MSE,Rab,Suzuki}. Also, theoretical investigations suggest that the Peierls stress is lower for a shuffle than for a glide dislocation \cite{PizzUNP}. Our 
results are then in agreement with experiments and with bulk calculations.

The influence of system size was checked by performing
additional simulations with 124 and 508 atoms. For the smaller size the elastic
limit was slightly larger, $-14.6$\%, whereas it remains at $-13.6$\% for the
larger size. These results are consistent with the trend obtained from
potential calculations, the elastic limit being larger for smaller
systems \cite{God04PRB}. However, this effect is very small and there were no
noticeable differences on the dislocation formation process, suggesting that
196 atoms are enough. The suitability of the minimal basis was checked by performing simulations
with the smaller system and the more accurate basis, leading to similar results.
We also investigated the influence of the relative
positions of the steps on the top and bottom surfaces. In figure~\ref{SZ196},
the two step edges belong to two distant shuffle planes, what results in two
opposite steps on the bottom surface after relaxation. We built a 128 atoms system  
with steps lined up in the same shuffle
plane. Under similar conditions, we found the same behavior than with the
124-atom system, though with a slightly larger elastic limit of $-15.6$\%. Thus,
both the effects of system size and step location appear to be negligible for the
process of dislocation formation.

In a previous study, using classical potentials, we had investigated the
influence of several parameters such as the step height and the orientation of
the applied stress \cite{God04PRB}. We found that the step height has only a
quantitative effect, a higher step lowering the elastic limit. Such an
investigation implies several simulations with larger systems, and is beyond
the scope of the present work. However, we expect this conclusion to remain
valid in our case. Regarding stress orientation, it has been shown that a
perfect 60$^\circ$ dislocation is nucleated for a wide range of stress
orientations, in agreement with a Schmid factor analysis \cite{God04PRB}.
However, an intriguing case is $\alpha=45^\circ$, for which the Schmid factors
are equal for screw and 60$^\circ$ dislocations. We have performed the
simulation in compression for this orientation, finding that plasticity occurs
at $-$20.2\%. The analysis of atomic displacements is difficult in this
case. Some atoms are displaced according to the formation of a 60$^\circ$, and
others according to the formation of a screw dislocation. However, we were unable to clearly
identify dislocations segments.

An important factor in the process of dislocation formation is the
surface and step termination. In fact, in compression, the formation of bonds
between the step atoms and the lower terrace leads to an easier nucleation,
starting from the step edge. In traction, dislocation formation also occurred
in the vicinity of the step. Therefore, surface and step termination are expected to play a key
role in realistic surfaces. We have then simulated a system of 128 Si atoms,
whose surfaces and steps were passivated with hydrogen atoms
(Fig.~\ref{sat_h}). In that case, the dimers of the p(2$\times1$) surface
reconstruction are symmetric. The system is elastically strained up to $-16.7$\%,
so that the shuffle plane passing through the step edge is increasingly
sheared. In that case, however, the hydrogen atoms prevent the formation of
bonds between step edge atoms and the terrace, and plasticity first occurred
inside the slab with the formation of a dislocation dipole, each dislocation
moving progressively towards one surface. Thus, the relaxation mechanism
clearly depends on the surface and step state, with or without hydrogen atoms.
However, it is noteworthy that the calculated elastic limits remain very
similar.

In this work, we have considered a (100) surface. For the cubic diamond
structure, another important surface orientation is (111). A simple Schmid factor
analysis suggests that larger applied stresses would be required for forming
dislocations. In addition, the same analysis suggests that a screw dislocation would be nucleated
for the (111) surface instead of a 60$^\circ$ dislocation.

In conclusion, we have investigated the process of dislocation formation from
surface steps in a stressed covalent solid, using first principles
calculations. Our study has been restricted to two-dimensional systems, in the context of athermal nucleation at 0K.
It has been shown that a simple surface defect, such as a step,
facilitates the formation of dislocations, by lowering the elastic limit and
initiating the plastic deformation. For the orientations considered, perfect
60$^\circ$ dislocations were nucleated, in agreement with a previous analysis
\cite{God04PRB}. It has also been shown that, with passivated steps and
surfaces, the formation mechanism is different but occurs for similar applied
strains. Overall, the elastic limits are reached for very large applied
stresses, much higher than expected in experiments. However, it is very likely that
thermal activation and higher defects will
considerably decrease these stresses, while preserving the formation
mechanisms. Performing {\em ab initio} finite temperature molecular dynamics
for these systems remains presently a huge task, but it is obviously the next
goal to attain in the future.

\section*{Acknowledgements}
The authors thank the Psi-K network for financial support of this work, and J.~Junquera
and E.~Anglada in the {\sc siesta} group for providing localized orbitals basis.





\begin{thebibliography}{44}
\expandafter\ifx\csname natexlab\endcsname\relax\def\natexlab#1{#1}\fi
\expandafter\ifx\csname bibnamefont\endcsname\relax
  \def\bibnamefont#1{#1}\fi
\expandafter\ifx\csname bibfnamefont\endcsname\relax
  \def\bibfnamefont#1{#1}\fi
\expandafter\ifx\csname citenamefont\endcsname\relax
  \def\citenamefont#1{#1}\fi
\expandafter\ifx\csname url\endcsname\relax
  \def\url#1{\texttt{#1}}\fi
\expandafter\ifx\csname urlprefix\endcsname\relax\def\urlprefix{URL }\fi
\providecommand{\bibinfo}[2]{#2}
\providecommand{\eprint}[2][]{\url{#2}}

\bibitem[{\citenamefont{Moulin et~al.}(1999)\citenamefont{Moulin, Condat, and
  Kubin}}]{Mou99PMA}
\bibinfo{author}{\bibfnamefont{A.}~\bibnamefont{Moulin}},
  \bibinfo{author}{\bibfnamefont{M.}~\bibnamefont{Condat}}, \bibnamefont{and}
  \bibinfo{author}{\bibfnamefont{L.~P.} \bibnamefont{Kubin}},
  \bibinfo{journal}{Phil. Mag. A} \textbf{\bibinfo{volume}{79}},
  \bibinfo{pages}{1995} (\bibinfo{year}{1999}).

\bibitem[{\citenamefont{Dunstan}(1997)}]{Dun97JMSME}
\bibinfo{author}{\bibfnamefont{J.}~\bibnamefont{Dunstan}}, \bibinfo{journal}{J.
  Mater. Sci.: Materials in Electronic} \textbf{\bibinfo{volume}{8}},
  \bibinfo{pages}{337} (\bibinfo{year}{1997}).

\bibitem[{\citenamefont{Jain et~al.}(1997)\citenamefont{Jain, Harker, and
  Cowley}}]{Jai97PMA}
\bibinfo{author}{\bibfnamefont{S.~C.} \bibnamefont{Jain}},
  \bibinfo{author}{\bibfnamefont{A.~H.} \bibnamefont{Harker}},
  \bibnamefont{and} \bibinfo{author}{\bibfnamefont{R.~A.}
  \bibnamefont{Cowley}}, \bibinfo{journal}{Phil. Mag. A}
  \textbf{\bibinfo{volume}{75}}, \bibinfo{pages}{1461} (\bibinfo{year}{1997}).

\bibitem[{\citenamefont{Wu and Weatherly}(2001)}]{Wu01PMA}
\bibinfo{author}{\bibfnamefont{R.~X.} \bibnamefont{Wu}} \bibnamefont{and}
  \bibinfo{author}{\bibfnamefont{G.~C.} \bibnamefont{Weatherly}},
  \bibinfo{journal}{Phil. Mag. A} \textbf{\bibinfo{volume}{81}},
  \bibinfo{pages}{1489} (\bibinfo{year}{2001}).

\bibitem[{\citenamefont{Wu and Xu}(1998)}]{Wu98PML}
\bibinfo{author}{\bibfnamefont{Y.~Q.} \bibnamefont{Wu}} \bibnamefont{and}
  \bibinfo{author}{\bibfnamefont{Y.~B.} \bibnamefont{Xu}},
  \bibinfo{journal}{Phil. Mag. Lett.} \textbf{\bibinfo{volume}{78}},
  \bibinfo{pages}{9} (\bibinfo{year}{1998}).

\bibitem[{\citenamefont{Scandian et~al.}(1999)\citenamefont{Scandian, Azzouzi,
  Maloufi, Michot, and George}}]{Sca99PSS}
\bibinfo{author}{\bibfnamefont{C.}~\bibnamefont{Scandian}},
  \bibinfo{author}{\bibfnamefont{H.}~\bibnamefont{Azzouzi}},
  \bibinfo{author}{\bibfnamefont{N.}~\bibnamefont{Maloufi}},
  \bibinfo{author}{\bibfnamefont{G.}~\bibnamefont{Michot}}, \bibnamefont{and}
  \bibinfo{author}{\bibfnamefont{A.}~\bibnamefont{George}},
  \bibinfo{journal}{Phys. Stat. Sol.\ A} \textbf{\bibinfo{volume}{171}},
  \bibinfo{pages}{67} (\bibinfo{year}{1999}).

\bibitem[{\citenamefont{Rabier et~al.}(2001)\citenamefont{Rabier, Cordier,
  Demenet, and Garem}}]{Rab01MSE}
\bibinfo{author}{\bibfnamefont{J.}~\bibnamefont{Rabier}},
  \bibinfo{author}{\bibfnamefont{P.}~\bibnamefont{Cordier}},
  \bibinfo{author}{\bibfnamefont{J.~L.} \bibnamefont{Demenet}},
  \bibnamefont{and} \bibinfo{author}{\bibfnamefont{H.}~\bibnamefont{Garem}},
  \bibinfo{journal}{Mater. Sci. Eng.\ A} \textbf{\bibinfo{volume}{A309-A310}},
  \bibinfo{pages}{74} (\bibinfo{year}{2001}).

\bibitem[{\citenamefont{Pirouz et~al.}(2001)\citenamefont{Pirouz, Demenet, and
  Hong}}]{Pir01PMA}
\bibinfo{author}{\bibfnamefont{P.}~\bibnamefont{Pirouz}},
  \bibinfo{author}{\bibfnamefont{J.~L.} \bibnamefont{Demenet}},
  \bibnamefont{and} \bibinfo{author}{\bibfnamefont{M.~H.} \bibnamefont{Hong}},
  \bibinfo{journal}{Phil. Mag. A} \textbf{\bibinfo{volume}{81}},
  \bibinfo{pages}{1207} (\bibinfo{year}{2001}).

\bibitem[{\citenamefont{Pirouz et~al.}(1999)\citenamefont{Pirouz, Samant, Hong,
  Moulin, and Kubin}}]{Pir99JMR}
\bibinfo{author}{\bibfnamefont{P.}~\bibnamefont{Pirouz}},
  \bibinfo{author}{\bibfnamefont{A.~V.} \bibnamefont{Samant}},
  \bibinfo{author}{\bibfnamefont{M.~H.} \bibnamefont{Hong}},
  \bibinfo{author}{\bibfnamefont{A.}~\bibnamefont{Moulin}}, \bibnamefont{and}
  \bibinfo{author}{\bibfnamefont{L.~P.} \bibnamefont{Kubin}},
  \bibinfo{journal}{J. Mar. Res.} \textbf{\bibinfo{volume}{14}},
  \bibinfo{pages}{2783} (\bibinfo{year}{1999}).

\bibitem[{\citenamefont{Frank}(1950)}]{Fra50Sym}
\bibinfo{author}{\bibfnamefont{F.~C.} \bibnamefont{Frank}}, in
  \emph{\bibinfo{booktitle}{Symposium on plastic deformation of crystalline
  solids}} (\bibinfo{publisher}{Carnegie Institute of Technology},
  \bibinfo{address}{Pittsburgh}, \bibinfo{year}{1950}), p.~\bibinfo{pages}{89}.

\bibitem[{\citenamefont{Kamat and Hirth}(1990)}]{Kam90JAP}
\bibinfo{author}{\bibfnamefont{S.~V.} \bibnamefont{Kamat}} \bibnamefont{and}
  \bibinfo{author}{\bibfnamefont{J.~P.} \bibnamefont{Hirth}},
  \bibinfo{journal}{J. Appl. Phys.} \textbf{\bibinfo{volume}{67}},
  \bibinfo{pages}{6844} (\bibinfo{year}{1990}).

\bibitem[{\citenamefont{Beltz and Freund}(1993)}]{Bel93PSS}
\bibinfo{author}{\bibfnamefont{G.~E.} \bibnamefont{Beltz}} \bibnamefont{and}
  \bibinfo{author}{\bibfnamefont{L.~B.} \bibnamefont{Freund}},
  \bibinfo{journal}{Phys. Stat. Sol.\ (b)} \textbf{\bibinfo{volume}{180}},
  \bibinfo{pages}{303} (\bibinfo{year}{1993}).

\bibitem[{\citenamefont{George and Michot}(1993)}]{Geo93MSE}
\bibinfo{author}{\bibfnamefont{A.}~\bibnamefont{George}} \bibnamefont{and}
  \bibinfo{author}{\bibfnamefont{G.}~\bibnamefont{Michot}},
  \bibinfo{journal}{Mater. Sci. Eng.} \textbf{\bibinfo{volume}{A164}},
  \bibinfo{pages}{118} (\bibinfo{year}{1993}).

\bibitem[{\citenamefont{Michot et~al.}(1999)\citenamefont{Michot, Azzouzi,
  Maloufi, de~Oliviera, Scandia, and George}}]{Mic99NATO}
\bibinfo{author}{\bibfnamefont{G.}~\bibnamefont{Michot}},
  \bibinfo{author}{\bibfnamefont{H.}~\bibnamefont{Azzouzi}},
  \bibinfo{author}{\bibfnamefont{N.}~\bibnamefont{Maloufi}},
  \bibinfo{author}{\bibfnamefont{M.~L.} \bibnamefont{de~Oliviera}},
  \bibinfo{author}{\bibfnamefont{C.}~\bibnamefont{Scandia}}, \bibnamefont{and}
  \bibinfo{author}{\bibfnamefont{A.}~\bibnamefont{George}}, in
  \emph{\bibinfo{booktitle}{Multiscale Phenomena in plasticity}}, edited by
  \bibinfo{editor}{\bibfnamefont{J.}~\bibnamefont{L\'epinoux}},
  \bibinfo{editor}{\bibfnamefont{D.}~\bibnamefont{Mazi\`ere}},
  \bibinfo{editor}{\bibfnamefont{V.}~\bibnamefont{Pontikis}}, \bibnamefont{and}
  \bibinfo{editor}{\bibfnamefont{G.}~\bibnamefont{Saada}}
  (\bibinfo{publisher}{Kluwer Academic Publishers},
  \bibinfo{address}{Dordrecht}, \bibinfo{year}{1999}), vol.
  \bibinfo{volume}{367} of \emph{\bibinfo{series}{NATO ASI Series E: Applied
  Sciences}}, p. \bibinfo{pages}{117}.

\bibitem[{\citenamefont{Gally and Argon}(2001)}]{Gal01PMA}
\bibinfo{author}{\bibfnamefont{B.~J.} \bibnamefont{Gally}} \bibnamefont{and}
  \bibinfo{author}{\bibfnamefont{A.~S.} \bibnamefont{Argon}},
  \bibinfo{journal}{Phil. Mag. A} \textbf{\bibinfo{volume}{81}},
  \bibinfo{pages}{699} (\bibinfo{year}{2001}).

\bibitem[{\citenamefont{Argon and Gally}(2001)}]{Arg01SM2}
\bibinfo{author}{\bibfnamefont{A.~S.} \bibnamefont{Argon}} \bibnamefont{and}
  \bibinfo{author}{\bibfnamefont{B.~J.} \bibnamefont{Gally}},
  \bibinfo{journal}{Scripta Materialia} \textbf{\bibinfo{volume}{45}},
  \bibinfo{pages}{1287} (\bibinfo{year}{2001}).

\bibitem[{\citenamefont{Marsh}(1963)}]{Mar63INT}
\bibinfo{author}{\bibfnamefont{D.~M.} \bibnamefont{Marsh}},
  \emph{\bibinfo{title}{Fracture of Solids}}
  (\bibinfo{publisher}{Interscience}, \bibinfo{year}{1963}), p.
  \bibinfo{pages}{119}.

\bibitem[{\citenamefont{Smith}(1968)}]{Smi68IJES}
\bibinfo{author}{\bibfnamefont{E.}~\bibnamefont{Smith}}, \bibinfo{journal}{Int.
  J. Engng. Sci.} \textbf{\bibinfo{volume}{6}}, \bibinfo{pages}{9}
  (\bibinfo{year}{1968}).

\bibitem[{\citenamefont{Cullis et~al.}(1996)\citenamefont{Cullis, Pidduck, and
  Emeny}}]{Cul96JCG}
\bibinfo{author}{\bibfnamefont{A.~G.} \bibnamefont{Cullis}},
  \bibinfo{author}{\bibfnamefont{A.~J.} \bibnamefont{Pidduck}},
  \bibnamefont{and} \bibinfo{author}{\bibfnamefont{M.~T.} \bibnamefont{Emeny}},
  \bibinfo{journal}{J. Cryst. Growth} \textbf{\bibinfo{volume}{158}},
  \bibinfo{pages}{15} (\bibinfo{year}{1996}).


\bibitem[{\citenamefont{Jagannadham and
  Narayan}(1991{\natexlab{a}})}]{Jag91}
\bibinfo{author}{\bibfnamefont{K.}~\bibnamefont{Jagannadham}} \bibnamefont{and}
  \bibinfo{author}{\bibfnamefont{J.}~\bibnamefont{Narayan}},
  \bibinfo{journal}{Mater. Sci. Eng.} \textbf{\bibinfo{volume}{B8}},
  \bibinfo{pages}{107} (\bibinfo{year}{1991}{\natexlab{a}});
  \bibinfo{journal}{J. Electron. Mater.} \textbf{\bibinfo{volume}{20}},
  \bibinfo{pages}{767} (\bibinfo{year}{1991}{\natexlab{b}}).

\bibitem[{\citenamefont{Beltz and Freund}(1995)}]{Bel95Sym}
\bibinfo{author}{\bibfnamefont{G.~E.} \bibnamefont{Beltz}} \bibnamefont{and}
  \bibinfo{author}{\bibfnamefont{L.~B.} \bibnamefont{Freund}}, in
  \emph{\bibinfo{booktitle}{Thin Films: Stresses and Mechanical Properties V.
  Symposium}}, edited by \bibinfo{editor}{\bibfnamefont{S.~P.}
  \bibnamefont{Baker}}, \bibinfo{editor}{\bibfnamefont{C.~A.}
  \bibnamefont{Ross}}, \bibinfo{editor}{\bibfnamefont{P.~H.}
  \bibnamefont{Townsend}}, \bibinfo{editor}{\bibfnamefont{C.~A.}
  \bibnamefont{Volkert}}, \bibnamefont{and}
  \bibinfo{editor}{\bibfnamefont{P.}~\bibnamefont{Borgesen.}}
  (\bibinfo{publisher}{Mater. Res. Soc}, \bibinfo{address}{Pittsburgh},
  \bibinfo{year}{1995}), p.~\bibinfo{pages}{93}.

\bibitem[{\citenamefont{Zou and Cockayne}(1996)}]{Zou96JAP}
\bibinfo{author}{\bibfnamefont{J.}~\bibnamefont{Zou}} \bibnamefont{and}
  \bibinfo{author}{\bibfnamefont{D.~J.~H.} \bibnamefont{Cockayne}},
  \bibinfo{journal}{J. Appl. Phys.} \textbf{\bibinfo{volume}{79}},
  \bibinfo{pages}{7632} (\bibinfo{year}{1996}).

\bibitem[{\citenamefont{Junqua and Grilhé}(1997)}]{Jun97PML}
\bibinfo{author}{\bibfnamefont{N.}~\bibnamefont{Junqua}} \bibnamefont{and}
  \bibinfo{author}{\bibfnamefont{J.}~\bibnamefont{Grilhé}},
  \bibinfo{journal}{Phil. Mag. Lett.} \textbf{\bibinfo{volume}{75}},
  \bibinfo{pages}{125} (\bibinfo{year}{1997}).

\bibitem[{\citenamefont{Nandedkar}(1993)}]{Nan93AMM}
\bibinfo{author}{\bibfnamefont{A.~S.} \bibnamefont{Nandedkar}},
  \bibinfo{journal}{Acta Metall. Mater.} \textbf{\bibinfo{volume}{41}},
  \bibinfo{pages}{3455} (\bibinfo{year}{1993}).

\bibitem[{\citenamefont{Ichimura and Narayan}(1995{\natexlab{a}})}]{Ich95}
\bibinfo{author}{\bibfnamefont{M.}~\bibnamefont{Ichimura}} \bibnamefont{and}
  \bibinfo{author}{\bibfnamefont{J.}~\bibnamefont{Narayan}},
  \bibinfo{journal}{Phil. Mag. A} \textbf{\bibinfo{volume}{72}},
  \bibinfo{pages}{297} (\bibinfo{year}{1995}{\natexlab{a}});
  \bibinfo{journal}{Mater. Sci. Eng.\ B} \textbf{\bibinfo{volume}{31}},
  \bibinfo{pages}{299} (\bibinfo{year}{1995}{\natexlab{b}}).

\bibitem[{\citenamefont{Juan et~al.}(1996)\citenamefont{Juan, Sun, and
  Kaxiras}}]{Jua96PML}
\bibinfo{author}{\bibfnamefont{Y.-M.} \bibnamefont{Juan}},
  \bibinfo{author}{\bibfnamefont{Y.}~\bibnamefont{Sun}}, \bibnamefont{and}
  \bibinfo{author}{\bibfnamefont{E.}~\bibnamefont{Kaxiras}},
  \bibinfo{journal}{Phil. Mag. Lett.} \textbf{\bibinfo{volume}{73}},
  \bibinfo{pages}{233} (\bibinfo{year}{1996}).

\bibitem[{\citenamefont{Zimmerman and Gao}(1996)}]{Zim96EMRS}
\bibinfo{author}{\bibfnamefont{J.~A.} \bibnamefont{Zimmerman}}
  \bibnamefont{and} \bibinfo{author}{\bibfnamefont{H.}~\bibnamefont{Gao}},
  \bibinfo{journal}{Mat. Res. Soc. Symp. Proc.} \textbf{\bibinfo{volume}{399}},
  \bibinfo{pages}{401} (\bibinfo{year}{1996}).

\bibitem[{\citenamefont{Aslanides and Pontikis}(1998)}]{Asl98PML}
\bibinfo{author}{\bibfnamefont{A.}~\bibnamefont{Aslanides}} \bibnamefont{and}
  \bibinfo{author}{\bibfnamefont{V.}~\bibnamefont{Pontikis}},
  \bibinfo{journal}{Phil. Mag. Lett.} \textbf{\bibinfo{volume}{78}},
  \bibinfo{pages}{377} (\bibinfo{year}{1998}).

\bibitem[{\citenamefont{Dong et~al.}(1998)\citenamefont{Dong, Schnitker, Smith,
  and Srolovitz}}]{Don98JAP}
\bibinfo{author}{\bibfnamefont{L.}~\bibnamefont{Dong}},
  \bibinfo{author}{\bibfnamefont{J.}~\bibnamefont{Schnitker}},
  \bibinfo{author}{\bibfnamefont{R.~W.} \bibnamefont{Smith}}, \bibnamefont{and}
  \bibinfo{author}{\bibfnamefont{D.~J.} \bibnamefont{Srolovitz}},
  \bibinfo{journal}{J. Appl. Phys.} \textbf{\bibinfo{volume}{83}},
  \bibinfo{pages}{217} (\bibinfo{year}{1998}).


\bibitem[{\citenamefont{Brochard
  et~al.}(2000{\natexlab{a}})\citenamefont{Brochard, Beauchamp, and
  Grilh\'e}}]{Bro00}
\bibinfo{author}{\bibfnamefont{S.}~\bibnamefont{Brochard}},
  \bibinfo{author}{\bibfnamefont{P.}~\bibnamefont{Beauchamp}},
  \bibnamefont{and} \bibinfo{author}{\bibfnamefont{J.}~\bibnamefont{Grilh\'e}},
  \bibinfo{journal}{Phil. Mag. A} \textbf{\bibinfo{volume}{80}},
  \bibinfo{pages}{503} (\bibinfo{year}{2000}{\natexlab{a}});
  \bibinfo{journal}{Phys. Rev. B} \textbf{\bibinfo{volume}{61}},
  \bibinfo{pages}{8707} (\bibinfo{year}{2000}{\natexlab{b}}).

\bibitem[{\citenamefont{Pizzagalli et~al.}(2003)\citenamefont{Pizzagalli,
  Beauchamp, and Rabier}}]{Piz03PMA}
\bibinfo{author}{\bibfnamefont{L.}~\bibnamefont{Pizzagalli}},
  \bibinfo{author}{\bibfnamefont{P.}~\bibnamefont{Beauchamp}},
  \bibnamefont{and} \bibinfo{author}{\bibfnamefont{J.}~\bibnamefont{Rabier}},
  \bibinfo{journal}{Phil. Mag. A} \textbf{\bibinfo{volume}{83}},
  \bibinfo{pages}{1191} (\bibinfo{year}{2003}).

\bibitem[{\citenamefont{Godet et~al.}(2003)\citenamefont{Godet, Pizzagalli,
  Brochard, and Beauchamp}}]{God03JPCM}
\bibinfo{author}{\bibfnamefont{J.}~\bibnamefont{Godet}},
  \bibinfo{author}{\bibfnamefont{L.}~\bibnamefont{Pizzagalli}},
  \bibinfo{author}{\bibfnamefont{S.}~\bibnamefont{Brochard}}, \bibnamefont{and}
  \bibinfo{author}{\bibfnamefont{P.}~\bibnamefont{Beauchamp}},
  \bibinfo{journal}{J. Phys.: Condens. Matter} \textbf{\bibinfo{volume}{15}},
  \bibinfo{pages}{6943} (\bibinfo{year}{2003}).

\bibitem[{\citenamefont{Troullier and Martins}(1991)}]{Tro91PRB}
\bibinfo{author}{\bibfnamefont{N.}~\bibnamefont{Troullier}} \bibnamefont{and}
  \bibinfo{author}{\bibfnamefont{J.~L.} \bibnamefont{Martins}},
  \bibinfo{journal}{Phys. Rev. B} \textbf{\bibinfo{volume}{43}},
  \bibinfo{pages}{1993} (\bibinfo{year}{1991}).

\bibitem[{\citenamefont{Ordej{\'o}n et~al.}(1996)\citenamefont{Ordej{\'o}n,
  Artacho, and Soler}}]{Ord96PRB}
\bibinfo{author}{\bibfnamefont{P.}~\bibnamefont{Ordej{\'o}n}},
  \bibinfo{author}{\bibfnamefont{E.}~\bibnamefont{Artacho}}, \bibnamefont{and}
  \bibinfo{author}{\bibfnamefont{J.~M.} \bibnamefont{Soler}},
  \bibinfo{journal}{Phys. Rev. B} \textbf{\bibinfo{volume}{53}},
  \bibinfo{pages}{R10441} (\bibinfo{year}{1996}).

\bibitem[{\citenamefont{Soler et~al.}(2002)\citenamefont{Soler, Artacho, Gale,
  Garc\'{\i}a, Junquera, Ordej\'on, and S\'anchez-Portal}}]{Sol02JPCM}
\bibinfo{author}{\bibfnamefont{J.~M.} \bibnamefont{Soler}},
  \bibinfo{author}{\bibfnamefont{E.}~\bibnamefont{Artacho}},
  \bibinfo{author}{\bibfnamefont{J.~D.} \bibnamefont{Gale}},
  \bibinfo{author}{\bibfnamefont{A.}~\bibnamefont{Garc\'{\i}a}},
  \bibinfo{author}{\bibfnamefont{J.}~\bibnamefont{Junquera}},
  \bibinfo{author}{\bibfnamefont{P.}~\bibnamefont{Ordej\'on}},
  \bibnamefont{and}
  \bibinfo{author}{\bibfnamefont{D.}~\bibnamefont{S\'anchez-Portal}},
  \bibinfo{journal}{J. Phys.: Condens. Matter} \textbf{\bibinfo{volume}{14}},
  \bibinfo{pages}{2745} (\bibinfo{year}{2002}).

\bibitem[{\citenamefont{Monkhorst and Pack}(1976)}]{Mon76PRB}
\bibinfo{author}{\bibfnamefont{H.~J.} \bibnamefont{Monkhorst}}
  \bibnamefont{and} \bibinfo{author}{\bibfnamefont{J.~D.} \bibnamefont{Pack}},
  \bibinfo{journal}{Phys. Rev. B} \textbf{\bibinfo{volume}{13}},
  \bibinfo{pages}{5188} (\bibinfo{year}{1976}).

\bibitem[{\citenamefont{Hirth and Lothe}(1982)}]{Hir82WIL}
\bibinfo{author}{\bibfnamefont{J.~P.} \bibnamefont{Hirth}} \bibnamefont{and}
  \bibinfo{author}{\bibfnamefont{J.}~\bibnamefont{Lothe}},
  \emph{\bibinfo{title}{Theory of dislocations}} (\bibinfo{publisher}{Wiley},
  \bibinfo{address}{New York}, \bibinfo{year}{1982}).

\bibitem[{\citenamefont{Godet et~al.}(2004)\citenamefont{Godet, Pizzagalli,
  Brochard, and Beauchamp}}]{God04PRB}
\bibinfo{author}{\bibfnamefont{J.}~\bibnamefont{Godet}},
  \bibinfo{author}{\bibfnamefont{L.}~\bibnamefont{Pizzagalli}},
  \bibinfo{author}{\bibfnamefont{S.}~\bibnamefont{Brochard}}, \bibnamefont{and}
  \bibinfo{author}{\bibfnamefont{P.}~\bibnamefont{Beauchamp}},
  \bibinfo{journal}{Phys. Rev. B} \textbf{\bibinfo{volume}{70}},
  \bibinfo{pages}{054109} (\bibinfo{year}{2004}).


\bibitem[{shu()}]{shuffleglide}
\bibinfo{note}{There are two inequivalent \{111\} planes in the diamond cubic
  structure, the widely spaced shuffle and the narrowly spaced glide
  \cite{Hir82WIL}.}

\bibitem[{\citenamefont{Chadi}(2002)}]{Cha87PRL}
\bibinfo{author}{\bibfnamefont{D.~J.}~\bibnamefont{Chadi}},
  \bibinfo{journal}{Phys. Rev. Lett.} \textbf{\bibinfo{volume}{59}},
  \bibinfo{pages}{1691} (\bibinfo{year}{1987}).


\bibitem[{\citenamefont{Rabier et~al.}(2000)\citenamefont{Rabier, Cordier,
  Tondellier, Demenet, and Garem}}]{Rab}
\bibinfo{author}{\bibfnamefont{J.}~\bibnamefont{Rabier}} {\it et al.},
  \bibinfo{journal}{J. Phys.: Condens. Matter} \textbf{\bibinfo{volume}{12}},
  \bibinfo{pages}{10059} (\bibinfo{year}{2000});
  \bibinfo{journal}{Scripta Materialia} \textbf{\bibinfo{volume}{45}},
  \bibinfo{pages}{1259} (\bibinfo{year}{2001}).

\bibitem[{\citenamefont{Suzuki et~al.}(2001)}]{Suzuki}
\bibinfo{author}{\bibfnamefont{T.}~\bibnamefont{Suzuki}} {\it et al.},
  \bibinfo{journal}{Phil. Mag. Lett.} \textbf{\bibinfo{volume}{77}},
  \bibinfo{pages}{173} (\bibinfo{year}{1998});
  \bibinfo{journal}{Phil. Mag. A} \textbf{\bibinfo{volume}{79}},
  \bibinfo{pages}{2637} (\bibinfo{year}{1999});

\bibitem[{\citenamefont{Pizzagalli and Beauchamp}(2001)}]{PizzUNP}
\bibinfo{author}{\bibfnamefont{L.} \bibnamefont{Pizzagalli}}
  \bibnamefont{and} \bibinfo{author}{\bibfnamefont{P.} \bibnamefont{Beauchamp}} 
  (\bibinfo{note}{unpublished}).

\bibitem[{\citenamefont{Godet at~al}(2002)}]{God02SM}
\bibinfo{author}{\bibfnamefont{J.}~\bibnamefont{Godet}},
  \bibinfo{author}{\bibfnamefont{L.}~\bibnamefont{Pizzagalli}},
  \bibinfo{author}{\bibfnamefont{S.}~\bibnamefont{Brochard}}, \bibnamefont{and}
  \bibinfo{author}{\bibfnamefont{P.}~\bibnamefont{Beauchamp}},
  \bibinfo{journal}{Scripta Materialia} \textbf{\bibinfo{volume}{47}},
  \bibinfo{pages}{481} (\bibinfo{year}{2002}).


\end{thebibliography}
\end{document}